Thermodynamics of interface in the freezing of colloidal suspensions: from macroscale to the microscale

Lilin Wang[1], Zhijun Wang[2*]

1-School of Materials Science and Engineering, Xi'an University of Technology, Xi'an 710048, P. R. China

2-State Key Laboratory of Solidification Processing, Northwestern Polytechnical University, Xi'an 710072, P. R. China

Abstract: A long controversy of ice lensing exists in the research of frost heave. By elucidating the mechanical and thermodynamics equilibria at the interface, we present the thermodynamics of the water/ice interface from macroscale to microscale for the freezing of colloidal suspensions. The application of the Clapeyron equation is confirmed both at macroscale to microscale via curvature effect. The thermodynamics at the interface indicates the initial of ice lensing/banding from the growth of pore ice, determined by the critical curvature undercooling instead of the critical fracture of frozen fringe. It is also proposed that the packing status of the porous structure in the particle layer ahead of the water/ice interface determines the ice lensing behaviors. The results presented here are different scenarios from previous investigations of freezing colloidal suspensions, and may shed light on the researches of this area.

    The frost heave is a popular and important phenomenon in cold region, responsible for the prime damage to the constructions [1, 2]. However, the mechanism of frost heave remains unsolved despite extensive prior researches. Many models have been proposed to illustrate the ice lensing behavior [3-6]. The frozen fringe is the crucial region to explain the growth of ice lens. Recent investigations suggest the ice

[*]Corresponding author. Tel.:86-29-88460650; fax: 86-29-88491484
E-mail address: zhjwang@nwpu.edu.cn (Zhijun Wang)

lens can grow even without frozen fringe[7, 8]. The particle constitutional supercooling theory has been also used to explain the ice-lens formation in the case with/without frozen fringe [9]. However, very recently, the quantitative measurement of the interface undercooling confirmed the limited contribution of the particle constitutional supercooling on the ice lensing [10-12]. These results indicate that the previous theoretical investigations of frozen fringe did not completely elucidate the ice lens growth. Based on the recent experimental work on ice lens formation, the growth of ice spear in the defect of particle packing ahead the solid/liquid interface may be one of the key procedures in the ice-lensing [13]. The instability of the water/ice interface will be the key clues in the theoretical analyses of ice lens formation.

In the ice lens growth, several aspects have been emphasized, the mechanics of ice and water in the porous media, the phase change during freezing, the water flow in the porous media during ice growth. All these factors affect the interface status in the ice propagation. The mechanics can alter the equilibrium melting point of the interface. The water flow supplies the water for ice growth and accompanies the depression of water pressure ahead the interface. The phase change of freezing will release the latent heat and bring the change of system volume. It should be noted that the ice lens was also observed in system with water replaced by other substances [2]. The freezing is a kind of crystal growth phenomenon, a typical Stefan problem with a free boundary for transfer equations. The boundary conditions at the moving interface are essential for the frost heave problem.

Quantifying the relationship between the water and ice at the interface in the colloidal suspensions has been a subject of interest for many years. The Clapeyron equation is the prior foundation in describing the change of melting temperature at the water/ice interface [14, 15]. However, there have been many argues on the application of Clapeyron equation in the freezing soil. The discrepancy from the experiments has been attributed to the irreversible thermodynamics of ice growth [16]. As a counterpart, according to the theory of solidification, the interface satisfies the assumption of local equilibrium generally and hence there is the Gibbs-Thomson (G-T) condition at the interface [17]. The curvature effect in the G-T condition is consistent with the Clapeyron equation. Accordingly, the Clapeyron equation is applicable in the freezing process during ice lensing. Some other reasons should be responsible for the inconsistency between the measurements and the theoretical predictions. The solute effect is the most likely candidate of depression of melting points.

At the interface, the conditions of thermodynamic equilibrium include the thermal, mechanical and chemical equilibria [18]. In this letter, we do not consider the impurities in the water and ice, and hence the osmotic pressure of dilute solution from van't Hoff's equation will not be involved in the pressure analyses during water flow. The thermal and mechanical equilibria at the interface will be the major concerns.

The experimental setup of Miller et al. [19] is the comprehensive model for analyzing the frost heave and ice lensing behavior, which is represented in Fig.1. The previous analyses mainly based on the generalized Clapeyron equation,

$$\frac{u_w}{\rho_w} - \frac{u_i}{\rho_i} = L\frac{T_I - T_f}{T_f} \qquad (1)$$

where $u_w$ and $u_i$ are the gauge pressures of the water and ice at the interface, $\rho_w$ and $\rho_i$ are the densities of water and ice, $T_I$ is the temperature at the interface, $T_f$ is the freezing point of water, $L$ is the latent heat of fusion at $T_f$. The application of this equation on different cases has been well discussed, mainly from the macroscopic view. The macroscopic analyses should involve the boundary conditions and the internal pressure variation. During the ice growth, the particles are extruded from the freezing ice and pushed by the interface to form a particle layer. In another view, the particle exchanges with water to supply water for ice growth and there is an internal local water flow. The driving force of the particle layer migration is the interface tension. Objectively, there is a pressure depression in the particle layer which is also called suction pressure. The pressure depression and the water flow is equivalent to the Darcy's flow,

$$q = -\frac{\kappa}{\mu}\nabla u_w \qquad (2)$$

where $q$ is the flux, $\nabla u_w$ is the pressure gradient, and $\mu$ is the viscosity. And the flux is equal to the growth velocity of the interface $v_I$,

$$q = v_I \qquad (3)$$

In the system of Fig.1, the boundary pressure condition of far field from the interface is $u_{wb}$ and $u_{ib}$. At the interface, the Eq. 1 will be

$$\frac{u_{wb} - \delta u_w}{\rho_w} - \frac{u_{ib}}{\rho_i} = L\frac{T_I - T_f}{T_f}, \qquad (4)$$

$\delta u_w$ is the depression of pressure in the water flow in the boundary layer. There is a restriction for the $\delta u_w$, $u_{wb} - \delta u_w > 0$.

The physical measurement of the suction pressure is around tens of kPa [14, 20], offering a undercooling around dozens of mK. It should be noted that in previous measurement of undercooling in open system without external load, the temperature was great larger than the theoretical prediction. However, the precise measurement of the interface undercooling was in the same order of the theoretical prediction by excluding the solute effect. The interface undercooling comes from the depression of pressure of water because of the dynamic Darcy's flow driving by the growth of interface, irrelevant to the particle induced thermodynamic osmotic pressure. When the system is with external force as that in the frost heave, the ice can has a burden of dozens of MPa of $u_{ib}$. In the porous media with water, the porous soil can afford the most burden from the ice while the fluid of water is determined by the external pressure $u_{wb}$. In this case, the great difference between the pressure of water and ice at the interface will lead to the depression of melting point for several degree centigrade. The external force has the dominant contribution on the interface thermodynamics compared with the Darcy's flow from ice growth. According to the constraint of $u_{wb} - \delta u_w > 0$, with increased external water pressure $u_{wb}$, there may be large suction pressure at the interface and the contribution of Darcy's flow may be increased.

Accordingly, the macroscopic mechanical equilibrium at the interface will be

$$u_{wb} + u_{pb} = u_{ib} \tag{5}$$

by ignoring the small force of accelerating or decelerating the particle layer ahead the interface, where $u_{pb}$ is the porous mechanical boundary conditions. In the local water flow region, the decrease of the pressure $\delta u_w$ converts into the increased pressure on the porous $\delta u_p$ by drag force of water flow $\delta u_w = \delta u_p$, and

$$u_{wb} - \delta u_w + u_{pb} + \delta u_p = u_{ib} \tag{6}.$$

Although the macroscale analyses show the thermal equilibria at the interface by analyzing the mechanical condition at the both side of the interface, it does not show the mechanical equilibrium at the interface in the consideration of the great difference between the pressures of ice and water at the interface. It is impossible for a large discrepancy of pressure between $u_w$ and $u_i$ during a near equilibrium planar growth of ice. Moreover, we do not know that whether there is an upper limit of the pressure difference between the ice and water at the interface. Therefore, there must be something at the interface to balance the pressure discrepancy at the difference sides of the interface. It is needed to zoom in the interface at the microscale. The microscale analysis is also helpful to discover the secrets of frost heave. We believe that the initial of frost heave is mainly related to the ice growth, while the mechanical behavior is subsequent consequence.

The interpretation of ice growth at the microscale is much different from that at the macroscale. The microscopic interface of the ice growth is illustrated in Fig. 2. The interaction model between a single particle and the interface has been well analyzed [21], where the particle is balanced by the disjoint force of surface tension and the viscous force of water flow. In the colloidal suspensions, the growth of the ice lenses can drive the migration of particle layer. For each particle, there is still a balance by the disjoint force of surface tension and the viscous force of water flow.

There is one choice to analyze the interface behind the particle with a premelted layer. In this region the premelting thermodynamics determines the equilibrium temperature. However, the situation will be much complex. Differently, the viscous force from $\delta u_w$ is much larger than that in the single particle model because it is the summation of the viscous force in the migrating particle layer. If $u_{pb}$ is not zero, the particle will endure much larger force by $u_{pb} + \delta u_p$. Moreover, the curvature effect indicates the pressure of liquid film in the premelting layer is much larger than that in the bulk ice. In this analyses, there will be unknown thickness of the premelting layer, $d$. Even we do not suspect the application of the Clapeyron equation and the premelting thermodynamics in this confined layer with extreme stress, we have one more unknown parameters $d$ and the interface temperature can not be determined by the Clapeyron equation unless there is another condition.

The better choice is to consider the curved interface between the particles which is also related the growth of pore ice in previous investigations. We will start the analysis from a very simple model. In the simple model as shown in Fig. 3, with two blocks pressed on the planar interface, the local stress of the ice will be increased, the melting temperature will decrease based on the Clapeyron equation or G-T relationship. In the linear temperature gradient, the interface will move backward to the cold region with a curve of grooving shape $z_I$ determined by,

$$\frac{u_w}{\rho_w} - \frac{u_i}{\rho_i} = L\frac{T_I - T_f}{T_f}$$
$$u_i - u_w = \frac{2\sigma_{iw}}{r(z_I)} \quad (7)$$
$$T_I = T_f - Gz_I$$

where $r$ is the radius at the interface. As the stress increases, the two separated grooves will join together. The central point of the curve will have a curvature pointing to the ice. The Clapeyron equation can be applied at this point, and the temperatures decreases. The curved interface balanced the stress difference between the ice and water. As the stress further increases, the curvature of the tip increases and the interface undercooling increases. There will be a critical situation for the tip stability. The critical curvature of the pore ice growth has been estimated in several models based on the geometrical analyses. Here the instability of the tip is analyzed based on the Clapeyron equation or G-T relationship. In the temperature gradient, the tip is warmest point at the curve. It indicates that the tip have the minimum curvature on the curve distribution around the tip. The curvature distribution is determined by the shape of the pore ice. On this requirement, there will be a maximum undercooling for a given space between the separated blocks. The solution is hidden in the Eq. 7.

Up to now, we can not give the analytical solution of the critical curvature even with mathematical software. However, we can present an example properly very close to the solution. The situation of Fig. 3 is much similar to the Staffman-Taylar finger growth, where the shape may be described by [22]

$$z - z_0 = \frac{b^2}{R\pi^2}\ln(\cos(\frac{\pi x}{b})) \quad (8)$$

where $z_0$ is the tip position, $R$ is the tip radius and $b = \lambda L$ is the effective finger width determined by the channel width $L$ and an effective parameter $\lambda \approx 0.9$ for smaller tip velocity. We have analyzed the curvature distribution along the finger during the cellular growth in directional solidification. There is a transition from the maximum to the local minimum for the tip curvature, determined by the shape characteristic $R/b \approx 0.39$. $b$ is almost constant for a given channel width $L$. In the pore ice growth, for smaller undercooling, there is smaller curvature for larger tip radius to satisfy the $R/b > 0.39$. However, as the tip undercooling increases, the tip curvature increases and $R/b < 0.39$. The tip will have the maximum curvature, and then the finger will grow into the warm region to decrease the tip temperature and hence to enlarge the tip radius to obey the Clapeyron equation and linear temperature gradient. Once the tip growth into the warm region, the local stress of the ice at the tip will decreases rapidly, and the tip is in the undercooled state based on the Clapeyron equation. The tip will grow faster. Finally, the two blocks with too large stress will be entrapped by the growth of pore ice. By the way, the root of the finger is determined by the thermodynamics of premelting layer.

Based on the microscopic analysis of interface thermodynamics, the curved interface balanced the interface temperature and the mechanical discrepancy on the two sides of the interface. With a particle layer, the local stress of these blocks on the ice is related to the macroscopic mechanical conditions. Actually, the analysis of the Clapeyron equation with macroscopic mechanical balance on the interface is untenable with a planar interface because there is no mechanical balance at the interface unless the interface is curved. It is the mechanical balance at the microscale providing the foundation of the Clapeyron equation, and the application of Clapeyron equation at the macroscale is the mean field assumption of the stress at the interface in the microscale. Therefore it is consistent for the curved interface in the microscale and the Clapeyron equation in the macroscale.

With the microscale analyses presented here, it is much easier to understand the existence of frozen fringe and ice banding formation mechanism. The pore ice growth determines the ice banding formation. Accordingly, the pore size is the crucial factor to be considered. In monodispersed colloidal suspensions, there will be no ice banding for particle with large size. On the one hand, the interface cannot move the particle. On the other hand, there will be no interface undercooling for the water in the large pore size or local stress to pushes the particle layer. As the particle size decreases, the interface can push the particle to form a particle boundary layer ahead the interface and the interface will be undercooled due to the curvature effect described in the Clapeyron equation and G-T relationship. The ice banding formation begins an ice spear from the pore ice growth, and then the lateral growth forms a new ice banding. The frozen fringe exists between two ice lenses where there are pore ice and the refusal water in the porous soil. In this case, the frozen fringe has no contribution to the ice banding formation, but a subsequent result of ice spear growth.

It should be noted that, the ploydispersity of the particle size is also a crucial factor affect the ice lenses. In the experiments of freezing soil, the size of the soil particle is dispersed in a large range. Accordingly, the porous structure of the particle layer ahead of the interface is hierarchical. During freezing, the water in the largest interconnected pores first freeze into pore ice composing the front edge of freezing. If the maximum of the pore size is large enough, the temperature of the front edge of ice is near the melting temperature of pure water and the so-called frozen fringe is left behind. In this case, the initial ice lens epitaxially grows from the existing pore ice. In this case, the ice banding forms without an ice spear.

In conclusion, we present the details of the application of Clapeyron equation on the ice lens growth from macroscale to microscale. It shows that the microscale representation of the Clapeyron equation is the foundation of the thermodynamics at the interface and its mean field case makes the the Clapeyron equation at the macroscale in the freezing colloidal suspensions. The macroscale analyses presents the boundary conditions of pressure at the interface, while the pore ice growth determines the maximum pressure deference between the water and ice, and hence the maximum of the interface undercooling. The microscale analyses indicate that the formation of frozen fringe and ice banding depend on the particle size distribution. For the monodispersed suspensions, the frozen fringe does not exist ahead of the ice lens, but left behind the lens in the cold region. The ice spear from the pore ice growth and the subsequent lateral growth are the intrinsic mechanism of ice lens growth. For the ploydispersed suspensions, the front edge of the frozen fringe is subjected to the pore ice growth in the larger connected pores and the ice banding forms in the fronzen fringe by lateral epitaxial growth of the existing pore ice. Anyway, the ice lens growth is determined by the pore ice growth related to the curvature effect and may be irrelevant to the related mechanical instability analyses of crack growth with Griffith fracture theory in the frozen fringe.

Acknowledgement: This research is supported by the Research Fund of the State Key Laboratory of Solidification Processing in NWPU, China (Grant Nos. 158-QP-2016 and SKLSP201627).

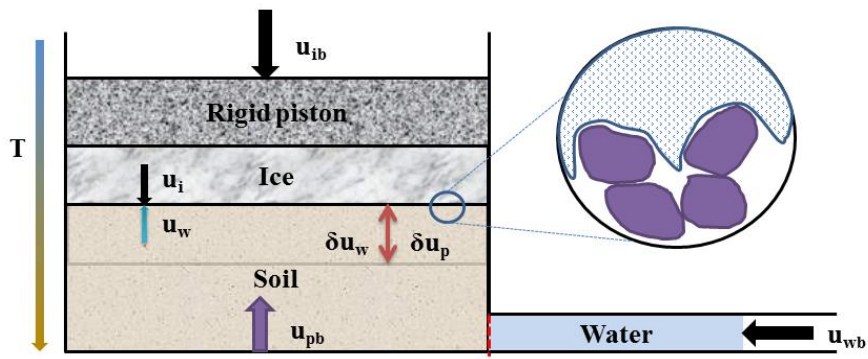

Figure 1 Representation of freezing soil experiments and the macroscale mechanical equilibrium. $u_w$ and $u_i$ are the gauge pressures of the water and ice at the interface, $u_{wb}$ and $u_{ib}$ are the boundary pressure conditions of far field from the interface, $\delta u_w$ is the depression of pressure in the water flow in the boundary layer, $\delta u_p$ is the increased pressure on the porous by drag force of water flow, $u_{pb}$ is the porous mechanical boundary conditions, T is the temperature.

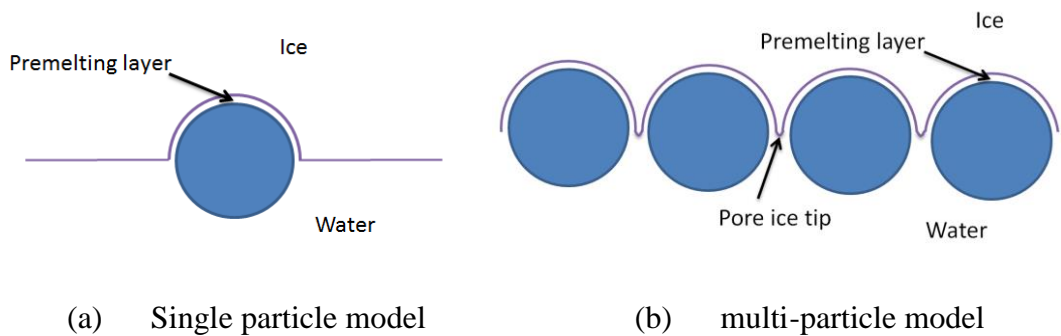

(a)　　Single particle model　　　　　(b)　　multi-particle model

Figure 2 The thermodynamic scenarios of the interface at the microscale during freezing of colloidal suspensions.

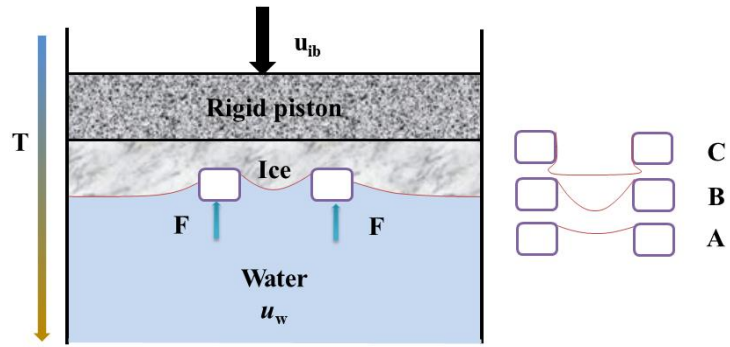

Figure 3 The sketch of simple model of pore ice growth, A, B and C show the cases with different pressures. $u_w$ is the gauge pressures of the water and ice at the interface, $u_{ib}$ is the boundary pressure conditions of far field from the interface, F is the force applied on the particles.